\documentstyle[twoside,fleqn,espcrc2,epsf]{article}

\title{A Simulation of the 't Hooft Model at finite $N_c$ with the Overlap
  Dirac Operator\thanks{Presented by F. Berruto.  Work supported
in part by grant DE-FG02-91ER40676 and by INFN.}}
\author{F.~Berruto, L.~Giusti, C.~Hoelbling, C.~Rebbi\\
[0.2cm]
Boston University - Department of Physics, 590 
Commonwealth Avenue, Boston MA 02215.}

\begin{document}

\begin{abstract}
We present some results of a numerical investigation of the 't Hooft model 
with 2,\ 3 and 4 colors, regularized on the lattice with overlap fermions.

\vspace{0mm}
\end{abstract}

\maketitle

Several years ago, in a pioneering investigation, 't~Hooft studied
$U(N_c)$ theories in the limit $N_c \to \infty$ with $C_t=g^2N_c$ kept
constant~\cite{thooft1}.  In two dimensions this led
him to a model of $QCD$ that exhibited important features of the full
theory.  Recently, the overlap formulation has made it possible to
introduce a lattice Dirac operator $D$ which in the massless limit preserves
a lattice form of chiral symmetry at finite lattice 
spacing~\cite{neuberger}.  This development, together with a renewed
interest in 't~Hooft's results, prompted us to perform a numerical
investigation of a class of two-dimensional non-Abelian models with
overlap lattice femions.  Precisely, we simulated models of
$QCD_2$ with $SU(N_c)$ and $U(N_c)$ color groups for $N_c=2,3,4$, in
the 't Hooft limit $C_t=g^2N_c=\mbox{const}$.  Since in two
dimensions only the $U(N_c)$ models exhibit a $U(1)$ axial anomaly, we
studied both classes of models in order to analyze the
behavior of the overlap operator in presence and absence of the axial
anomaly. Our systems are small enough that it is possible to
compute exactly in the overlap formulation the propagator $D^{-1}$ and
$\det (D)$, following the scheme used in~\cite{rebbi1}.  We
computed the masses of the pions as a function of the quark mass $m$
and, for both classes of models, we found the expected functional
dependence on $m$ and universality in $N_c$.  In the $U(N_c)$ case the
meson spectrum exhibits also a flavor singlet particle ($\eta '$),
whose mass is known analytically, and which we computed by following
the approximate method of Ref.~\cite{hamber} and by using the
Witten-Veneziano relation.  The details of our simulation, as well as
additional results that we could not include here because of lack of
space, will be presented in a separate article~\cite{berruto}.

\section{The 't Hooft Model with Overlap Fermions}

The 't Hooft model action in the overlap regularization reads
\begin{eqnarray}
S&=&\beta\sum_{x,\mu<\nu}\left[1-\frac{1}{2N_c}\mbox{Re}\ \mbox{Tr}\ U_{\mu
    \nu}(x) \right]\\
&+&\sum_{x,y}
\overline{\psi}(x)\left[ (1-\frac{ma}{2})D_{xy}+m\ \delta_{xy}
\right]   \psi(y)\nonumber
\label{laction}
\end{eqnarray}
where $U_{\mu \nu}$ is the usual Wilson plaquette, $a$ is the lattice
spacing and $\psi$ are 
$N_f$ fermion fields with mass $m$ and $\beta=2N_c/(ag)^2$. 
$D=(1+V)/a$ is the massless Neuberger-Dirac operator,  $V$ being the unitary 
component of $D_W-1/a$, where $D_W$ is the usual Wilson-Dirac operator. 

For $m=0$ the action 
(\ref{laction}) is invariant under a continuous symmetry 
$\delta \psi=\gamma_5(1-aD)\psi,\quad \delta
\overline{\psi}=\overline{\psi}\gamma_5$, {\it i.e.} 
a form of chiral symmetry that holds at
finite lattice spacing. 
The $N_f=2$ models exhibit an isotriplet of particles ($\pi$), 
whose masses vary with the quark
mass as~\cite{thooft1}~\cite{hamer}
\begin{eqnarray}
M_{\pi}^2&=&2\ \sqrt{\frac{C_t\  \pi}{3}}\ m+\ldots\ ,\ \ \mbox{$U(N_c)$
  models}
\label{mpuvsmq}\\
M_{\pi}^2&=&\frac{9}{\pi}\
(2^7C_t)^{\frac{1}{3}}(\frac{e^{\gamma}}{\pi})^{\frac{4}{3}}\
m^{\frac{4}{3}}+\ldots\ , \ \ \mbox{$SU(N_c)$
  models}\quad\quad \label{mpsvsmq}
\end{eqnarray}
Eq.(\ref{mpuvsmq}) has been derived for $N_c\rightarrow \infty$ while 
Eq.(\ref{mpsvsmq}) also involves a semiclassical WKB approximation.

In the $U(N_c)$ models in the chiral limit $M_{\eta '}^2=N_f\ g^2 / \pi$. 
The Witten-Veneziano formula for the $U(N_c)$ models gives
\begin{equation}
M_{\eta '}^2=\frac{4N_f}{f_{\pi}^2}\ \frac{\langle Q^2\rangle}{V} 
\label{wvr}
\end{equation}     

\section{The meson spectrum}

We performed extensive simulations with the exact Neuberger-Dirac operator 
on a $18\times 18$ lattice at $C_t=4/3$, {\it i.e.} 
$\beta=6,\ 13.5,\ 24$ for $N_c=2,3,4$ respectively. We computed $M_{\pi}$ for
$m=0.04,0.05,0.06,0.07,0.08,0.1$ so that the relation $N_xM_{\pi}> 4$
holds. The effects of
dynamical fermions have been included by weighting the observables
with the appropriate power of the fermion determinant.
The smallness of the lattice warrants this procedure (cfr.~\cite{rebbi1}),
which would lead to unacceptable variance on larger systems.

We computed the meson propagators projected over zero momentum 
from the fermion propagators and extracted the meson masses in a standard
manner. The errors have been estimated with the jacknife analysis. 

In table (\ref{mp}) we report the pion masses for 
the $U(N_c)$ and $SU(N_c)$ models with two flavors of dynamical fermions 
and in the quenched approximation.
\begin{table}[!htb]
\begin{center}
\begin{tabular}{||l|l|l|l||}
\hline\hline
\multicolumn{4}{||c||}{$U(N_c)$ models}\\ 
\hline
                 & $N_c=2$          & $N_c=3$          & $N_c=4$          \\
$m_q/g\sqrt{N_c}$&$M_{\pi}^2/g^2N_c$&$M_{\pi}^2/g^2N_c$&$M_{\pi}^2/g^2N_c$\\
\hline
\multicolumn{4}{||c||}{$N_f=2$}  \\ 
\hline
0.0346 & 0.045(10) & 0.047(2) & 0.055(6) \\ 
0.0433 & 0.056(9)  & 0.059(2) & 0.067(5) \\ 
0.0519 & 0.068(8)  & 0.072(3) & 0.079(4) \\ 
0.0606 & 0.080(7)  & 0.085(3) & 0.093(3) \\ 
0.0692 & 0.094(5)  & 0.098(3) & 0.107(2) \\ 
0.0866 & 0.124(4)  & 0.128(3) & 0.138(1) \\  
\hline
\multicolumn{4}{||c||}{quenched ($N_f=0$)} \\
\hline
0.0346 &   0.069(1)   & 0.063(1)  &  0.057(1)   \\ 
0.0433 &   0.083(1)   & 0.076(1)  &  0.071(1)   \\ 
0.0519 &   0.098(1)   & 0.091(1)  &  0.085(1)   \\ 
0.0606 &   0.113(1)   & 0.106(1)  &  0.100(1)   \\ 
0.0692 &   0.128(1)   & 0.121(1)  &  0.116(1)   \\ 
0.0866 &   0.161(1)   & 0.153(1)  &  0.148(1)   \\
\hline
\hline
\multicolumn{4}{||c||}{$SU(N_c)$ models}\\ 
\hline
                 & $N_c=2$          & $N_c=3$          & $N_c=4$          \\
$(m_q/gN_c^{\frac{1}{2}})^{2/3}$&$M_{\pi}/gN_c^{\frac{1}{2}}
$&$M_{\pi}/gN_c^{\frac{1}{2}}$&$M_{\pi}/gN_c^{\frac{1}{2}}$\\
\hline
\multicolumn{4}{||c||}{$N_f=2$} \\
\hline
0.1062  &  0.183(14)  &  0.206(23)   & 0.181(28)    \\ 
0.1233  &  0.211(12)  &  0.237(23)   & 0.211(26)    \\ 
0.1392  &  0.236(10)  &  0.263(21)   & 0.239(24)    \\ 
0.1543  &  0.258(8)   &  0.286(19)   & 0.266(22)    \\ 
0.1686  &  0.279(7)   &  0.307(16)   & 0.292(21)    \\ 
0.1957  &  0.321(5)   &  0.348(10)   & 0.341(17)    \\  
\hline
\multicolumn{4}{||c||}{quenched ($N_f=0$)} \\
\hline
0.1062  & 0.200(2)  & 0.212(1)   & 0.217(1)  \\ 
0.1233  & 0.227(2)  & 0.241(1)   & 0.246(1)  \\ 
0.1392  & 0.252(2)  & 0.267(1)   & 0.273(1)  \\ 
0.1543  & 0.276(1)  & 0.292(1)   & 0.298(1)  \\ 
0.1686  & 0.299(1)  & 0.316(1)   & 0.322(1)  \\ 
0.1957  & 0.343(1)  & 0.361(1)   & 0.367(1)  \\  
\hline\hline
\end{tabular}
\caption{\label{mp}}
\end{center}

\vskip -1.5cm

\end{table}
The quenched results get closer to 
the unquenched two flavors results when
$N_c$ gets larger, as one would expect since the large $N_c$ limit 
corresponds to discarding the internal quark loops. 
For the $U(N_c)$ case Eq.(\ref{mpuvsmq}) motivated us to do the fit 
\begin{equation}
\frac{M_{\pi}^2}{g^2N_c}=A+B\ \frac{m}{g\sqrt{N_c}}
\label{ufitp}
\end{equation}
We obtained $A=-0.012(17), \; B=1.57(19)$ for $N_c=2$, $A=-0.006(4), \;
B=1.53(4)$ for $N_c=3$ and $A=-0.006(8),\; B=1.66(8)$ for
$N_c=4$.

\begin{figure}[htb]
\epsfxsize=\hsize
\epsfbox{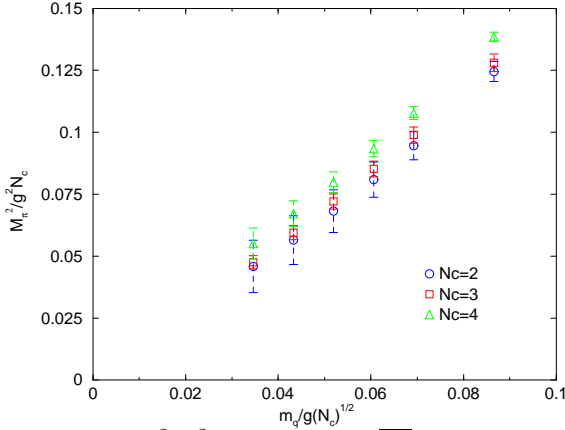}
\vspace{-12mm}
\caption{\it{$M^2_\pi/g^2N_c$ vs.~$m_q/g\sqrt{N_c}$ for $N_f=2$, $U(N_c)$ 
models}}
\vspace{-7mm}
\label{figpnf2u}
\end{figure}

\begin{figure}[htb]
\epsfxsize=\hsize
\epsfbox{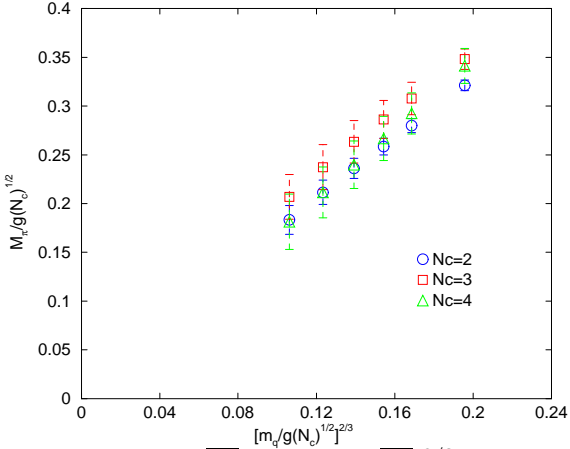}
\vspace{-12mm}
\caption{\it{$M_\pi/g\sqrt{N_c}$ vs.~$(m_q/g\sqrt{N_c})^{2/3}$ for $N_f=2$, $SU(N_c)$ models}}
\vspace{-7mm}
\label{figpnf2s}
\end{figure}

For the $SU(N_c)$ models, following Eq.(\ref{mpsvsmq}), 
we performed the fit
\begin{equation}
\frac{M_{\pi}}{g\sqrt{N_c}}=A+B\ (\frac{m}{g\sqrt{N_c}})^{\frac{2}{3}}
\label{sfitp}
\end{equation}
We obtained $A=0.022(26),\; B=1.524(120)$ for $N_c=2$, $A=0.045(43), \;
B=1.548(166)$ for $N_c=3$ and $A=-0.009(40), \; B=1.788(11)$ for
$N_c=4$. In the $SU(N_c)$ case we performed also the fit 
$M_{\pi}/g\sqrt{N_c}=C(m/g\sqrt{N_c})^{\gamma}$ and we
obtained $\gamma$ compatible with $2/3$ for $N_c=2,3,4$. 
Figures(\ref{figpnf2u},\ref{figpnf2s}) give evidence of universality 
in $N_c$ for the pion masses.

\begin{table}[!ht]
\begin{center}
\begin{tabular}{||l|l|l||}
\hline\hline
 $N_c=2$          & $N_c=3$          & $N_c=4$          \\
$M_{\eta '}^2/g^2N_c$&$M_{\eta '}^2/g^2N_c$&$M_{\eta '}^2/g^2N_c$\\
\hline
\multicolumn{3}{||c||}{Analytic} \\
\hline
0.159    & 0.106  & 0.079     \\
\hline
\multicolumn{3}{||c||}{Numerical from Eq.(\ref{etarebbi})} \\
\hline
0.159(8)   & 0.111(7)     & 0.084(8)     \\
\hline
\multicolumn{3}{||c||}{Numerical from Eq.(\ref{wvr})} \\
\hline
0.129(6)&0.097(7) &0.076(6) \\
\hline\hline
\end{tabular}
\caption{\label{metacl}}
\end{center}

\vskip -1.5cm

\end{table}

In the $U(N_c)$ models we computed also $M_{\eta '}$.  
The $\eta '$ propagator is given by
the difference between the connected and disconnected terms in the correlator
and, because of cancellations, the errors are larger than in the $\pi$ case.  
Therefore, we resorted the method proposed in~\cite{hamber} to compute
the $\eta '$ mass. The method exploits the quenched two-loop disconnected 
$\Gamma_q^{2-\mbox{loop}}$ and quenched one-loop connected 
$\Gamma_q^{1-\mbox{loop}}$ contributions to the $\eta '$ propagator:
\begin{equation}
M_{\eta '}^2=2M_{\pi}\lim_{t\rightarrow\infty}
\frac{\Gamma_q^{2-\mbox{loop}}(t)}
{|t|\Gamma_q^{1-\mbox{loop}}(t)}
\label{etarebbi}
\end{equation}
As one can see from table (\ref{metacl}) $M^2_{\eta '}$ 
computed using Eq.(\ref{etarebbi}) and $N_f=1$ 
compares remarkably well with the analytic result. 

An alternative calculation of $M_{\eta '}$ can be done by 
using the Witten-Veneziano formula.  
In the quenched approximation we computed
$f_{\pi}=0.774(3),\ 0.956(3),\ 1.115(5)$ for $N_c=2,3,4$ respectively. 
Since in our calculations we diagonalize the full Neuberger-Dirac operator, 
we could determine the number of zero modes, and thus the topological
charge, for each individual gauge configuration.  
We thus obtained $\langle Q^2
\rangle/V=0.0258(16),\ 0.0298(23),\ 0.0319(37)$ for $N_c=2,3,4$ respectively. 
(Notice that with increasing $N_c$ these values get closer and closer to 
the analytic value $\langle Q^2 \rangle/V=C_t/4\pi^2=0.0337$. 
Building on the trade-off between spatial and internal degrees of 
freedom that prompted the introduction of the 
$N_c\rightarrow \infty$ single plaquette model~\cite{eguchi}, one could
heuristically argue that the theory should be less 
affected by finite size effects increasing $N_c$. 
A rigorous justification of this
argument, however, would require a detailed analysis of discretization and
finite size effects that goes beyond the scope of our investigation.)
From $f_{\pi}$ and $\langle Q^2\rangle/V$, we use Eq.(\ref{wvr})
to calculate $M_{\eta'}$. 
The results, also in table (\ref{metacl}), are consistent for $N_c=3$ 
and 4 with those obtained with the method of Ref.~\cite{hamber} 
and with the analytical formula.

\end{document}